\documentclass[journal]{IEEEtran}
\usepackage{multirow,tabularx,caption,subcaption,bm,enumitem,kantlipsum,xcolor,hyperref,fontawesome5,balance,soul,graphicx,amsmath,arydshln,xcolor,booktabs}
\usepackage[table]{xcolor}
\ifCLASSINFOpdf

\else

\fi

\begin{document}
\title{CROSSAN: Towards Efficient and Effective Adaptation of Multiple Multimodal Foundation Models for Sequential Recommendation}

\author{Junchen Fu, Yongxin Ni, Joemon M. Jose, Ioannis Arapakis, Kaiwen Zheng,  Youhua Li,
        and  Xuri Ge% <-this % stops a space
\thanks{ 
J. Fu, K. Zheng, and J. Jose are with the School of Computing Science, University of Glasgow, Glasgow,
UK (e-mail: j.fu.3@research.gla.ac.uk, k.zheng.1@research.gla.ac.uk, and joemon.jose@glasgow.ac.uk). 
Y. Ni is with Boston University, Boston, United States (email: niyongxin2016@gmail.com)
I. Arapakis is with Telefonica Scientific Research, Barcelona, Spain (email: arapakis.ioannis@gmail.com).
Y. Li is with City University of Hong Kong, Hong Kong, China. (e-mail: youhuali2-c@my.cityu.edu.hk) 
X. Ge is with the School of Artificial Intelligence at Shandong University, Jinan, Shandong, China. (e-mail: xuri.ge@sdu.edu.cn) 
 (\textit{Corresponding author: Yongxin Ni and Xuri Ge.)}}
}

\markboth{Journal of \LaTeX\ Class Files,~Vol.~14, No.~8, August~2015}%
{Shell \MakeLowercase{\textit{et al.}}: Bare Demo of IEEEtran.cls for IEEE Journals}

\maketitle

\begin{abstract}
In this paper, we explore a less-studied yet practically important problem: how to efficiently and effectively adapt multiple ($>$2) multimodal foundation models (MFMs) for the sequential recommendation task. To this end, we propose a plug-and-play \underline{Cros}s-modal \underline{S}ide \underline{A}dapter \underline{N}etwork (\textbf{CROSSAN}), which leverages a fully decoupled side adapter-based paradigm to achieve efficient and scalable adaptation. Compared to the state-of-the-art efficient approaches, CROSSAN reduces training time by over 30\%, GPU memory consumption by 20\%, and trainable parameters by over 57\%, while enabling effective cross-modal learning across diverse modalities. To further enhance multimodal fusion, we introduce the \underline{M}ixture \underline{o}f \underline{M}odality \underline{E}xpert \underline{F}usion (\textbf{MOMEF}) mechanism. Extensive experiments on public benchmarks demonstrate that CROSSAN consistently outperforms existing methods, achieving 6.7\%--8.1\% performance improvements when adapting four foundation models with raw modalities. Moreover, the overall performance continues to improve as more MFMs are incorporated. We will release our code and datasets to faciliate future research.
\end{abstract}

\begin{IEEEkeywords}
Recommender Systems, Multimodal Foundation Models, Efficient
Adaptation, Sequential Recommendation, CROSSAN, MOMEF
\end{IEEEkeywords}

% \fi
\IEEEpeerreviewmaketitle

% \includegraphics{figs/grabs.pdf}


% \maketitle
\section{Introduction}
Multimodal Foundation Models (MFMs) have advanced rapidly, with models like ViT~\cite{dosovitskiy2020image}, BERT~\cite{devlin2018bert}, GPT~\cite{brown2020language}, VideoMAE~\cite{tong2022videomae}, and AST~\cite{gong2021ast} demonstrating exceptional performance in representing a wide range of raw modalities. At the same time, the increasing availability of recommendation datasets \cite{ni2023content} containing raw multimodal data (e.g., images, text, video, audio, etc.) provides a natural avenue for exploring how these powerful models can be effectively adapted for sequential recommendation tasks.

One intuitive approach that preserves enough information, is to adapt these models to the raw modalities of recommendation datasets \cite{yuan2023go,li2023multi,wu2021empowering}. Adaptation methods for MFMs that leverage raw modality information, such as fine-tuning and parameter-efficient fine-tuning (PEFT), are generally recognized for their ability to achieve better performance compared to traditional feature-based approaches \cite{yuan2023go,li2023multi,liu2023multimodal,ni2023content}. However, these approaches have been largely sidelined due to the central challenge of the significant computational costs associated with existing adaptation methods for multiple MFMs. Both full fine-tuning and PEFT techniques, such as Adapters \cite{houlsby2019parameter,fu2024exploring} and LoRA \cite{hu2021lora}, become increasingly expensive as the number of modality encoders grow\footnote{Recommender systems are typically retrained on a daily or weekly basis \cite{zhang2020retrain}, making costly training paradigms impractical in practice.}. This computational bottleneck makes it more challenging to investigate the adaptation of additional MFMs, leaving the potential of these models for sequential recommendation largely unexplored. 

Recently, the side adapter paradigm \cite{fu2024iisan, sung2022lst, xu2023side} has garnered significant attention for its superior efficiency in adaptation compared to traditional adapter-based or LoRA-based approaches. This efficiency stems from its fully decoupled design, which eliminates the need for gigantic computational graphs. However, existing studies on side adapters have primarily focused on single or dual-modal scenarios, leaving their potential for scalability to additional modalities and the integration of multiple multimodal foundation models (MFMs) unexplored. This gap highlights the need for further research to extend their applicability to more complex and diverse multimodal setups.

Motivated by the aforementioned challenges, this study aims to explore the under-explored area of adapting multiple ($>$2) multimodal foundation models (MFMs) for sequential recommendation and to propose a solution that enables efficient and effective adaptation. To this end, we introduce the \underline{Cros}s-modal \underline{S}ide \underline{A}dapter \underline{N}etwork (CROSSAN), a novel plug-and-play approach that addresses the computational inefficiencies of existing adaptation methods for multiple MFMs in item representation.  
To enhance multimodal interactions and improve overall effectiveness, we propose a cross-modal side adapter network, building on key insights from our preliminary study (\autoref{sec:preliminary_study}).
In our analysis, we demonstrate that cross-modal interaction enhances mutual information compared with independent side adapters. Moreover, in contrast to existing methods that fuse item representations through a fully connected layer \cite{fu2024iisan}, we adopt the Mixture of Modality Expert Fusion (MOMEF). This approach incorporates a fine-grained gating mechanism that enables the adaptive integration of input modalities, providing a straightforward yet highly effective solution for capturing intricate multimodal interactions and enhancing representation fidelity.
CROSSAN offers a scalable, efficient, and effective solution, achieving superior performance on public datasets. It outperforms existing efficient adaptation approaches, with its performance improvement becoming even more significant as more MFMs are integrated. Our contributions are listed below:
\begin{itemize}
    \item To achieve enhanced multimodal representation learning while maintaining high efficiency, we introduce CROSSAN, a simple yet effective, cross-modality side adapter method. 
    \item Building upon traditional concatenation-based fusion for multimodal item representation, we explore different fusion strategies and show that the Mixture of Modality Expert Fusion (MOMEF) mechanism effectively integrates item representations across modalities, leading to improved recommendation performance.
    \item Through extensive experiments, we demonstrate that the CROSSAN delivers superior performance and efficiency on the public dataset. These results highlight the potential of leveraging multiple MFMs with raw modalities for the multimodal sequential recommendation, paving the way for future research in this direction.
\end{itemize}
% \vspace{-0.1in}

\begin{figure}
  \centering
   \includegraphics[width=\linewidth]{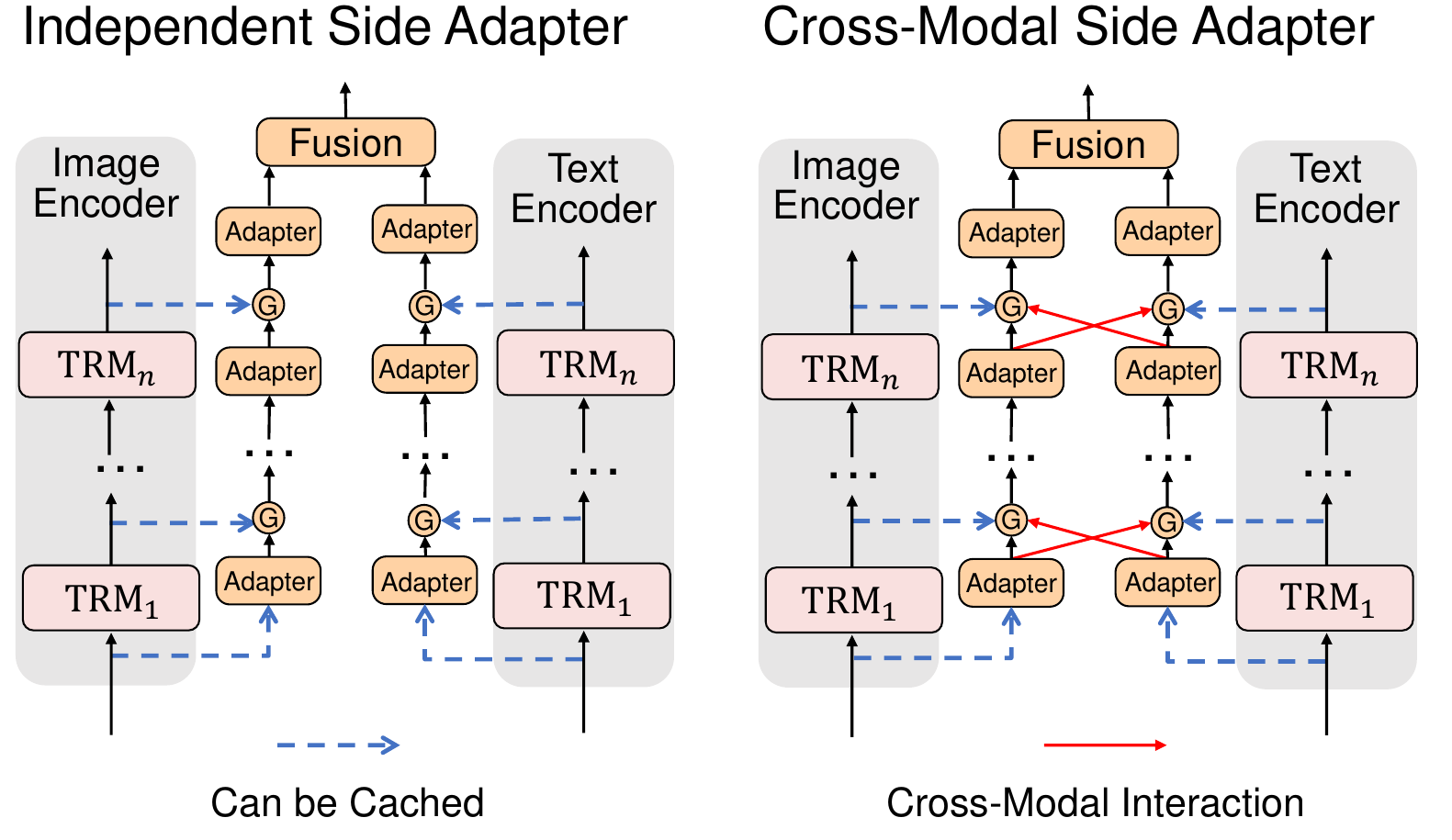}
  \caption{Independent vs. Cross-modal Side Adapter. During training, only the adapters and fusion layer are updated, while the rest of the foundation models remain frozen. 
  }
   % \vspace{-0.2in}
    \label{fig:independent_vs_cross} 
\end{figure}

\section{Preliminary study: Cross or Independent?}\label{sec:preliminary_study}

To maximize efficiency, we adopt a recently advanced, fully decoupled, side adapter paradigm with a caching strategy \cite{fu2024iisan, sung2022lst, xu2023side}, where the adapters are positioned outside the transformer models. This paradigm, while extensively studied for single-modality adaptation, remains underexplored in the context of multiple multimodal foundation models (MFMs). Therefore, we investigate two potential approaches: implementing the adapters either in a cross-modal configuration or independently, as illustrated in \autoref{fig:independent_vs_cross}. Preliminary experiments conducted on two commonly used combined modalities (text and image), using the Microlens-100K dataset, indicate that the cross-modal approach outperforms the independent method across four evaluation metrics of Hit ratio and NDCG (\autoref{tab:crossan_vs_indsan}). To understand this performance improvement, we further provide a theoretical analysis in \autoref{sec:theory}.

\begin{table}[h]
\centering
\caption{Performance comparison between INDSAN (Independent Side Adapter) and CROSSAN (Cross-Modal Side Adapter). ``*” denotes that the improvements are significant at the level of 0.05 with a paired T-test.}
\label{tab:crossan_vs_indsan}
\begin{tabular}{lcc}
\toprule
Metric & INDSAN & CROSSAN \\
\midrule
HR@10   & 0.0957  & \textbf{0.0970$^{*}$} \\
HR@20   & 0.1373 & \textbf{0.1393$^{*}$} \\
NDCG@10 & 0.0521  & \textbf{0.0537$^{*}$} \\
NDCG@20 & 0.0626  & \textbf{0.0644$^{*}$} \\
\hline
\hline
Mutual Information&0.0129 & \textbf{0.2001$^{*}$} \\
\bottomrule
\end{tabular}
% \vspace{-0.1in}
\end{table}

\section{Theoretical Analysis}
\label{sec:theory}
In this section, we provide a theoretical analysis demonstrating why the proposed Cross-Modal Side Adapter (CROSSAN) outperforms the Independent Side Adapter (INDSAN). For clarity, we focus on a representative scenario involving text and image modalities.

\subsection{Mutual Information as an Evaluation Metric}

Let $Z^{(v)}$ and $Z^{(t)}$ denote the learned visual and textual representations at a given network layer. The mutual information (MI)~\cite{paninski2003estimation} between these representations is defined based on entropy as:
\begin{equation}
I(Z^{(v)}; Z^{(t)}) = H(Z^{(v)}) - H(Z^{(v)} \mid Z^{(t)}),
\end{equation}
where $H(\cdot)$ denotes the entropy, which measures the uncertainty or information content of a random variable. Specifically, $H(Z^{(v)})$ quantifies the total uncertainty of the visual representations, while $H(Z^{(v)} \mid Z^{(t)})$ captures the remaining uncertainty in $Z^{(v)}$ after observing the textual representations $Z^{(t)}$.

A higher mutual information $I(Z^{(v)}; Z^{(t)})$ implies that the textual representations $Z^{(t)}$ provide more information about the visual representations $Z^{(v)}$, thus reducing their uncertainty. This reflects stronger dependency and more effective cross-modal integration, indicating that the learned representations capture complementary and aligned information across modalities, which is desirable for downstream tasks.  Many studies show that maximizing mutual information enhances representation quality \cite{guo2019learning, zhou2020s3, liao2021multimodal}. Thus, we use it as the evaluation metric for multimodal learning.

\subsection{Independent vs. Cross-Modal Adapters for Item Representation}

Based on the definition of mutual information in terms of entropy~\cite{paninski2003estimation}, where $H(\cdot)$ denotes entropy, the \textbf{INDSAN} model produces approximately independent modality-specific representations, denoted as $Z^{(v)}$ and $Z^{(t)}$. Specifically, we assume that the joint distribution of these representations satisfies $p\bigl(Z^{(v)}, Z^{(t)}\bigr) \approx p\bigl(Z^{(v)}\bigr)p\bigl(Z^{(t)}\bigr)$, indicating near-independence between modalities. This near-independence implies that one modality provides minimal information about the other, leading to $H\bigl(Z^{(v)} \mid Z^{(t)}\bigr) \approx H\bigl(Z^{(v)}\bigr)$ and, consequently, negligible mutual information:
\[
I_{\mathrm{INDSAN}}\bigl(Z^{(v)}, Z^{(t)}\bigr) = H\bigl(Z^{(v)}\bigr) - H\bigl(Z^{(v)} \mid Z^{(t)}\bigr) \approx 0.
\]
In contrast, \textbf{CROSSAN} establishes dependency through cross-modal interactions at each layer, where
\[
Z_{l}^{(v)} = f_{l}^{(v)}\bigl(X, Z_{l-1}^{(t)}\bigr), \quad Z_{l}^{(t)} = g_{l}^{(t)}\bigl(Y, Z_{l-1}^{(v)}\bigr).
\]
This ensures that $p\bigl(Z_{l}^{(v)}, Z_{l}^{(t)}\bigr) \neq p\bigl(Z_{l}^{(v)}\bigr)p\bigl(Z_{l}^{(t)}\bigr)$, meaning the uncertainty of the final representation $Z_{L}^{(v)}$ is reduced when $Z_{L}^{(t)}$ is known, i.e., $H\bigl(Z_{L}^{(v)} \mid Z_{L}^{(t)}\bigr) < H\bigl(Z_{L}^{(v)}\bigr)$. Consequently, the mutual information in CROSSAN is inherently positive:
\[
I_{\mathrm{CROSSAN}}\bigl(Z_{L}^{(v)}, Z_{L}^{(t)}\bigr) = H\bigl(Z_{L}^{(v)}\bigr) - H\bigl(Z_{L}^{(v)} \mid Z_{L}^{(t)}\bigr) \geq 0.
\]
This yields an upper bound mutual information than INDSAN. This continuous cross-layer interaction aligns with prior findings, indicating that deeper, multi-level interactions enhance shared information across modalities \cite{wu2021mutual, gu2022cross, yang2022vision}.

To empirically validate our theoretical analysis, we evaluate the final outputs of the textual and visual adapters using the optimal checkpoints for both INDSAN and CROSSAN, processing all items from the MicroLens-100K dataset. The results corroborate our conclusion, demonstrating that CROSSAN achieves significantly higher mutual information than INDSAN. Specifically, as illustrated in \autoref{tab:crossan_vs_indsan}, CROSSAN yields mutual information values more than 15 times greater than those observed with INDSAN. This notable difference supports the previous analysis and provides insights into the underlying reasons for CROSSAN’s performance improvement.\footnote{In the following section, we primarily adopt the state-of-the-art IISAN as the main comparison counterpart. In terms of mutual information, IISAN is theoretically equivalent to INDSAN due to the absence of dependencies between side adapters. The measured mutual information of IISAN is 0.0127, which is approximately the same as that of INDSAN. Compared to INDSAN, IISAN introduces an additional set of independent inter-modal side adapters that capture static multimodal information. However, dependencies among the side adapters are still absent.}

\begin{figure*}
  \centering
   \includegraphics[width=\linewidth]{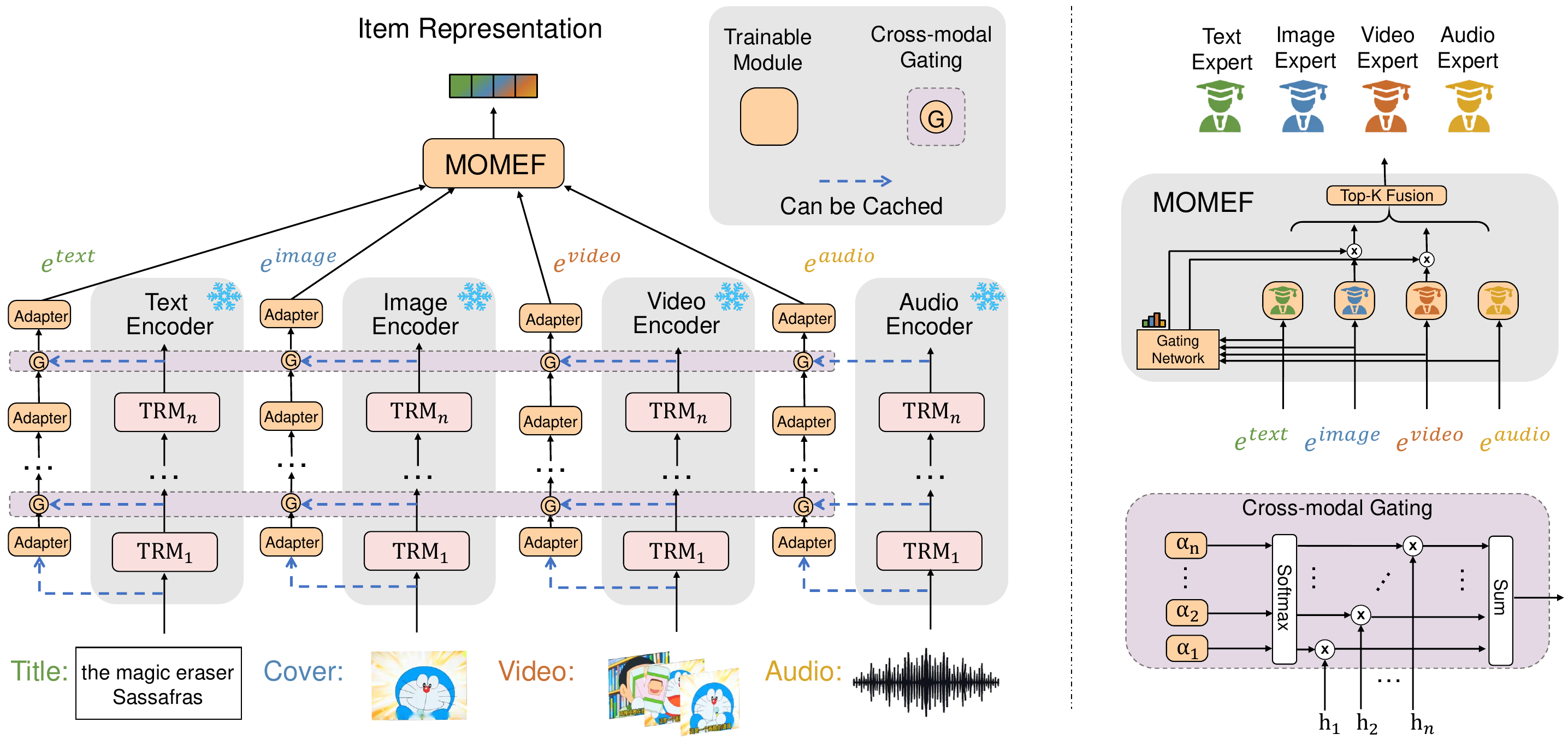}
  \caption{CROSSAN Overview. The example utilizes four MFMs, each paired with its respective side adapters for adaptation. A cross-modal gating mechanism is applied, combining the outputs of each modality's side adapter at the same layer-level with gated fusion, ensuring effective interaction across modalities. MOMEF (Mixture of Modality Expert Fusion) treats each input modality as an expert. The gating network assigns probabilities to each expert, selecting the top-k experts based on these probabilities for further processing.
  }
  % \vspace{-0.1in}
    \label{fig:cross_framework} 
\end{figure*}

\section{Methodology}
In this work, we introduce CROSSAN, a plug-and-play approach designed to adapt multiple MFMs. CROSSAN is designed to provide effective multimodal representations in a scalable and efficient manner, offering a general solution for the adaptation of multiple MFMs. The overview of CROSSAN is shown in \autoref{fig:cross_framework}. Our key innovation lies in designing a novel fully cross-modal gating between each modality's side adapters, dedicated to learning rich mutual information between modalities. Furthermore, to enable more effective multimodal fusion, we propose a Mixture of Modality Expert Fusion (MOMEF) network, which dynamically combines the multimodal outputs of all towers to achieve fine-grained fusion for each item.

\noindent \textbf{Problem Formulation.} Given a recommendation dataset $\mathcal{D} = \{\mathcal{U}, \mathcal{V}\}$, where $\mathcal{U}$ and $\mathcal{V}$ represent the set of users and items respectively, our objective in a multimodal sequential recommendation task is to predict the next item that a user $u$ will interact with, based on their past $n$ interactions. For multimodal recommendation, each item $v$ can have representations from $M$ different modalities, such as text ($v^{text}$), image ($v^{image}$), video ($v^{video}$), and audio ($v^{audio}$). Although additional modalities can be incorporated, depending on the application, in this paper, we mainly focus on these four standard modalities. Following \cite{fu2024iisan,yuan2023go}, we process each of the modalities using their corresponding pre-trained foundational models, such as BERT \cite{devlin2018bert} for text, ViT \cite{dosovitskiy2020image} for images, VideoMAE \cite{tong2022videomae} for videos, and Audio Spectrogram Transformer (AST) \cite{gong2021ast} for audio. By leveraging these pre-trained MFMs backbones, we obtain the hidden states for each modality (e.g., ${h_i^{text}}$, ${h_i^{image}}$, ${h_i^{video}}$, and ${h_i^{audio}}$) from their transformer layers (${TRM_i}$).

\noindent \textbf{Cross-modal Gating.} 
We employ a simple yet effective cross-modal gating mechanism, which differs from the traditional cross-attention mechanism proposed in \cite{huang2019ccnet}. Unlike cross-attention, which relies on computationally intensive attention mechanisms, our approach is based on straightforward tunable weights, offering a lightweight and efficient alternative. Specifically, for each modality at the $i$-th layer, the input to the adapter combines the output of the adapter from the $(i-1)$-th layer with the hidden states from the corresponding MFM. For instance, taking the side adapter's text modality at the $i$-th layer as an example, we define its input as follows:
\begin{equation}
    h^{t}_i = Adapter^{t}_{i} \left( \sum_{m \in M} \alpha^{m}_{i} h^{m}_{i-1} + \alpha^{t}_{i} h^{BERT}_{i-1} \right)
\end{equation}
where $\sum_{m \in M} \alpha^m_{i} + \alpha^{t}_{i} = 1$.

Here, $M$ represents the set of modalities considered in this work, specifically $M = \{text, image, video, audio\}$, and $t$ denotes the text modality. The parameter \(\alpha\) controls the learnable weight for each layer. We utilize the adapter block design proposed by \cite{houlsby2019parameter}, as it has proven to be highly effective in sequential recommendation tasks \cite{fu2024exploring,fu2024iisan}.

\noindent \textbf{Mixture-of-Modality Expert Fusion (MOMEF).} Furthermore, to enhance multimodal representation fusion while maintaining efficiency, we adopt a MOMEF method, drawing inspiration from the mixture-of-experts paradigm \cite{zhou2022mixture,dai2024deepseekmoe}, which dynamically combines the multimodal outputs of all towers to achieve a fine-grained fusion for each item. Specifically, each modality $m \in M$ has its own expert network producing an output vector $f_m(i)$ for item $i$. We compute importance scores $w_m(i)$ using a gating network based on fully connected layers and select the top-$k$ modalities $\text{TopK}(i)$ with the highest scores for each item. The final multimodal item representation $e^{final}$ is then obtained by dynamically weighting and combining the outputs of these top-$k$ modality experts:
\begin{equation}
    e^{final} = \sum_{m \in \text{TopK}(i)} w_m(i) \cdot f_m(i)
\end{equation}
where $\sum_{m \in \text{TopK}(i)} w_m(i) = 1$. This approach allows MOMEF to focus on the most pertinent modalities output for each item, leading to a more effective and fine-grained multimodal fusion.

Subsequently, the vector $e^{final}$ is fed into the sequential encoders to compute the final predicted score $\hat{y}_{ui}$ for user $u$ and item $i$, which is calculated as the product of the sequential encoder output and the corresponding item embedding. Note that our entire framework maintains high efficiency, as all trainable modules are primarily composed of linear layers and gating mechanisms, without relying on attention mechanisms.

\noindent \textbf{Loss Function.} Regarding training, we adopt the commonly used in-batch debiased Cross-Entropy loss function $\mathcal{L}_{CE}$~\cite{yi2019sampling,yuan2023go,ji2023online,fu2024iisan,ni2023content}, which is defined as:
\begin{equation}
D_{ui} = \exp(\hat{y}_{ui} - \log(p_i)) + \sum_{j \in [B], j \notin I_u} \exp(\hat{y}_{uj} - \log(p_j))
\end{equation}
\begin{equation}\label{eq:lce}
\mathcal{L}_{CE}=-\sum_{u \in \mathcal{U}} \sum_{i \in [2,...,n+1]} \log \frac{\exp(\hat{y}_{ui} - \log(p_i))}{D_{ui}}
\end{equation}
where $p_i$ represents the popularity of item $i$, $I_u$ denotes the set of items interacted by user $u$, and $B$ is the batch size. The item $n+1$ refers to the predicted item for user $u$.

\section{Experiment Setup}

\noindent \textbf{Dataset.} To assess the effectiveness of adapting multiple MFMs for recommendation tasks, we consider datasets that contain more than two raw modalities. Specifically, we use the publicly available Microlens-100K and the Microlens-50K datasets provided in \cite{ni2023content}.\footnote{To the best of our knowledge, Microlens is the only publicly available dataset that includes three or more raw modalities (e.g., text, images, video, and audio). We leave the exploration of additional datasets for future work as and when more modality-rich datasets become available.} The statistical details of the dataset are presented in \autoref{tab:dataset_micro}. Further evaluations on datasets with only two modalities are provided in the \autoref{sec:ap_more_eval}.

\noindent \textbf{Evaluation.}
Based on previous studies \cite{ni2023content, yuan2023go, fu2024exploring, fu2024iisan}, our approach implements a leave-one-out evaluation strategy. Specifically, the final item in the interaction sequence is set aside for testing, the second-to-last item is used for validation, and the rest of the sequence is employed for training. To evaluate our model's performance, we consider the HR (Hit Ratio) and NDCG (Normalized Discounted Cumulative Gain) metrics, which are aligned with previous studies \cite{ni2023content, yuan2023go, fu2024exploring}. Unless otherwise suggested, all reported results correspond to the test set. We also note that the predicted item is evaluated against the entire set of items \cite{krichene2020sampled}.

\noindent \textbf{Implementation Details.} We employ "bert-base-uncased", "vit-base-patch16-224", "MIT/ast-finetuned-audioset-10-10-0.4593", and "MCG-NJU/videomae-base" from Huggingface\footnote{\url{https://huggingface.co/}} as the text, image, audio, and video encoders, respectively.\footnote{The exploration of using LLMs as encoders is beyond the scope of this paper, primarily due to the challenge of managing asymmetry across multiple MFMs. However, we provide a preliminary investigation of advanced LLM and LVM models in the \autoref{sec:ap_more_eval}.} Our choice is informed by previous research in the field \cite{ni2023content,fu2024iisan,yuan2023go,jain2022hugging,dong2024musechat}. For video processing, we use the first three seconds of footage and extract 16 frames for VideoMAE, following its original setup \cite{tong2022videomae}, with corresponding audio processed by AST. For side adapters, we employ LayerDrop, dropping half the layers of each foundation model for efficiency \cite{fu2024iisan,sung2022lst}. 
We utilize a transformer-based sequential encoder to model user sequences, following the approach outlined in \cite{hou2022learning, yuan2023go}. The hidden dimension of the sequential encoder is set to 64 after a search in \{32, 64, 128\}, with two Transformer blocks and attention heads following \cite{fu2024exploring,fu2024iisan}.  The learning rate is optimized between 1e-5 and 1e-3, keeping dropout at 0.1 \cite{ni2023content}. We search batch sizes from 32 to 1024, selecting the largest based on GPU memory limits. Adapter hidden dimensions and LoRA ranks are tuned between 32 and 8192. Hyperparameters are determined by tuning on validation data, and all results are reported on the test set. All experiments are completed on an A6000 GPU.

\begin{table}
  \caption{Dataset Description.}
  \label{tab:dataset_micro}
  \renewcommand\tabcolsep{0.6pt}
  \scalebox{1}{
  \begin{tabular}{cccccc}
    \toprule
    \multirow{2}{*}{Dataset}&\multirow{2}{*}{Users}&\multirow{2}{*}{Items}&\multirow{2}{*}{Interaction}&\multirow{2}{*}{Raw Content}\\
    % &&&&&\\
    % Dataset & Users & Items & Interaction & Raw Content\\
    &&&&&\\
    \midrule
    Microlens-100K&100,000&19,738&719,405&Text\&Image\&Video\&Audio\\
    Microlens-50K&50,000&19,099&339,511&Text\&Image\&Video\&Audio\\
  \bottomrule
\end{tabular}
}
\end{table}

\section{Experiment}
Our evaluation addresses the following research questions: 
\begin{itemize}
    \item \textbf{RQ1:} How effective is CROSSAN compared with existing adaptation approaches, and does adapting more MFMs improve its performance compared to state-of-the-art efficient adaptation approaches?
    \item \textbf{RQ2:} How does CROSSAN's efficiency compare to state-of-the-art adaptation approaches?
    \item \textbf{RQ3}: How does each component affect the overall performance?
    \item \textbf{RQ4}: How does the hyperparameter affect CROSSAN?
\end{itemize}

\begin{table*}[htbp]
\centering
\caption{Performance comparison of CROSSAN on Microlens-100K and Microlens-50K with four types of raw modalities (\underline{T}ext, \underline{I}mage, \underline{V}ideo and \underline{A}udio, denoted as T, I, V, and A, respectively). ``*'' indicates that the improvements of the best models compared with previous state-of-the-art methods are significant at the level of 0.05 with paired T-test. 'Relative Improvement' is computed in comparison to the corresponding IISAN and its extension.}
\label{tab:microlens_merged}
\renewcommand{\arraystretch}{1}
\setlength{\tabcolsep}{6pt}
\begin{tabular}{lcccccccc}
\toprule
\multicolumn{1}{c}{} & \multicolumn{4}{c}{Microlens-100K} & \multicolumn{4}{c}{Microlens-50K} \\
\cmidrule(lr){2-5} \cmidrule(lr){6-9}
\multicolumn{1}{c}{Model} & HR@10 & NDCG@10 & HR@20 & NDCG@20 & HR@10 & NDCG@10 & HR@20 & NDCG@20 \\
\midrule
Full Finetuning (T+I)       & 0.0934 & 0.0499 & 0.1363 & 0.0607 & 0.0772 & 0.0408 & 0.1129 & 0.0498 \\
Adapter (T+I)\cite{houlsby2019parameter}   & 0.0962 & 0.0514 & 0.1376 & 0.0618 & 0.0765 & 0.0407 & 0.1132 & 0.0500 \\
LoRA (T+I)\cite{hu2021lora}               & 0.0866 & 0.0462 & 0.1298 & 0.0571 & 0.0644 & 0.0331 & 0.0975 & 0.0415 \\
\midrule
IISAN (T+I)\cite{fu2024iisan}             & 0.0960 & 0.0526 & 0.1366 & 0.0628 & 0.0771 & 0.0421 & 0.1121 & 0.0509 \\
IISAN-E (T+I+A)                           & 0.0953 & 0.0523 & 0.1353 & 0.0623 & 0.0777 & 0.0422 & 0.1137 & 0.0513 \\
IISAN-E (T+I+V)                           & 0.0939 & 0.0517 & 0.1341 & 0.0619 & 0.0775 & 0.0428 & 0.1150 & 0.0522 \\
IISAN-E (T+I+V+A)                         & 0.0949 & 0.0524 & 0.1350 & 0.0625 & 0.0790 & 0.0430 & 0.1135 & 0.0517 \\
\rowcolor{gray!20}
CROSSAN (T+I) \textbf{(ours)}             & 0.0999$^{*}$ & 0.0553$^{*}$ & 0.1428$^{*}$ & 0.0661$^{*}$ & 0.0806$^{*}$ & 0.0431$^{*}$ & 0.1177$^{*}$ & 0.0524$^{*}$ \\
\rowcolor{gray!20}
CROSSAN (T+I+A) \textbf{(ours)}          & 0.1012$^{*}$ & 0.0557$^{*}$ & 0.1445$^{*}$ & 0.0666$^{*}$ & 0.0811$^{*}$ & 0.0444$^{*}$ & 0.1188$^{*}$ & 0.0539$^{*}$ \\
\rowcolor{gray!20}
CROSSAN (T+I+V) \textbf{(ours)}          & 0.1006$^{*}$ & 0.0557$^{*}$ & 0.1428$^{*}$ & 0.0663$^{*}$ & 0.0808$^{*}$ & 0.0437$^{*}$ & 0.1183$^{*}$ & 0.0531$^{*}$ \\
\rowcolor{gray!20}
CROSSAN (T+I+V+A) \textbf{(ours)}        & \textbf{0.1033$^{*}$} & \textbf{0.0568$^{*}$} & \textbf{0.1452$^{*}$} & \textbf{0.0673$^{*}$} & \textbf{0.0847$^{*}$} & \textbf{0.0462$^{*}$} & \textbf{0.1222$^{*}$} & \textbf{0.0557$^{*}$} \\
\midrule
\textbf{Relative Improvement} \\
Text+Image                  & +3.86\% & +4.99\% & +4.43\% & +5.03\% & +4.25\% & +2.24\% & +4.81\% & +2.99\% \\
Text+Image+Audio            & +5.87\% & +6.17\% & +6.39\% & +6.43\% & +4.17\% & +4.95\% & +4.34\% & +4.88\% \\
Text+Image+Video            & +6.70\% & +7.18\% & +6.04\% & +6.74\% & +4.06\% & +2.04\% & +2.74\% & +1.75\% \\
Text+Image+Video+Audio      & +8.10\% & +7.70\% & +7.00\% & +7.15\% & +6.73\% & +6.97\% & +7.13\% & +7.21\% \\
\bottomrule
\end{tabular}
\end{table*}

\subsection{Effectiveness Evaluation (RQ1)}
We compare our approach against the popular efficient adaptation method Full finetuning, Adapter \cite{houlsby2019parameter}, LoRA \cite{hu2021lora}, and the state-of-the-art IISAN \cite{fu2024iisan}, as well as its intuitive extension to support additional modalities. IISAN and IISAN-E serve as our primary multimodal adaptation baselines for two reasons: (1) with two modalities, IISAN achieves competitive performance compared to other methods, and (2) our limited GPU memory (48GB) prevents us from using full fine-tuning or adapter-based approaches on more than two modalities, given their already high requirements for just two (see \autoref{tab:eff_compare}). 
Given that the original IISAN implementation is limited to two modalities, we extend its structure by incorporating additional intra-SAN layers to handle more multimodal foundation models (MFMs). In this expanded IISAN-E, we use a gated sum to combine the hidden states of all MFMs into the inter-SAN.

As shown in \autoref{tab:microlens_merged}, CROSSAN achieves the best performance on the Microlens-100k, outperforming all other adaptation approaches. Furthermore, we observe a progressive improvement in performance as more modalities from MFMs are incorporated. This trend becomes especially evident when compared to IISAN and its extension, IISAN-E, where the relative improvement increases with the introduction of additional modalities, as adding more MFMs to IISAN-E does not consistently result in better performance. This highlights the scalability and efficacy of CROSSAN over existing state-of-the-art efficient adaptation methods. To further validate these findings, we evaluate CROSSAN on the Microlens-50K dataset (\autoref{tab:microlens_merged}). The relative improvements remain consistent, with one exception: the T+I+V configuration shows comparable improvement to T+I. However, the overall trend of increasing performance with more MFMs and raw modalities persists. These results reaffirm the superiority of CROSSAN.

\begin{figure}[htbp]
  \centering
   \includegraphics[width=\linewidth]{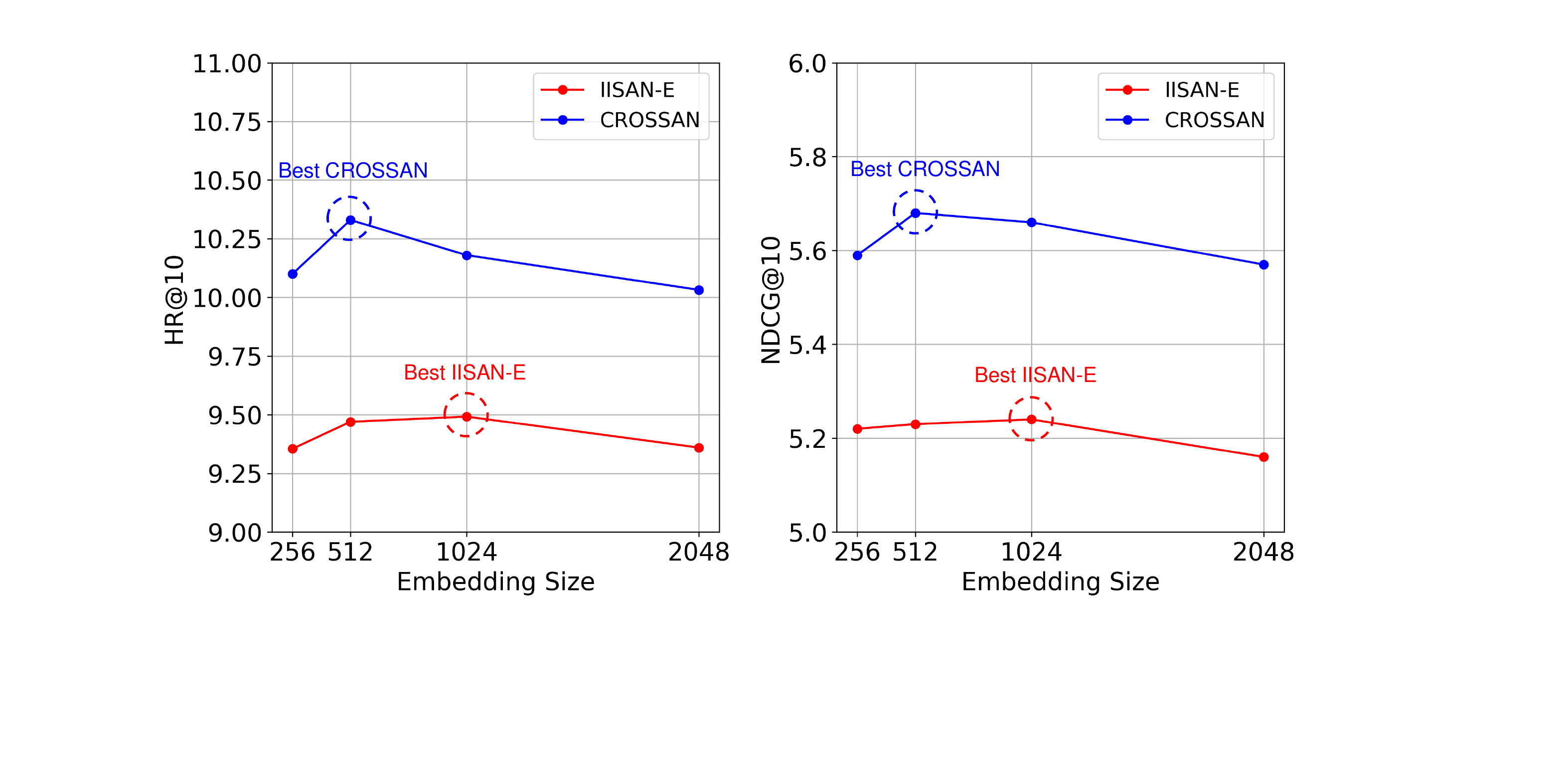}
  \caption{Optimal Embedding dimension for IISAN-E and CROSSAN
  }
    \label{fig:emb_independent_vs_cross}
\end{figure}

To further compare CROSSAN with additional state-of-the-art multimodal sequential recommendation models (MMSRs) beyond IISAN, we also include a recent state-of-the-art HM4SR model in our evaluation, as summarized in \autoref{tab:sota_msrs}. Across all evaluation metrics, CROSSAN consistently outperforms these strong baselines, demonstrating its superior effectiveness. These results further highlight the advantages of adapting raw modalities through multimodal foundation models (MFMs).

\begin{table}[htbp]
\centering
\caption{Performance comparison for state-of-the-art multimodal sequential recommendation approaches MicroLens datasets with respect to Hit Ratio (HR@K), and NDCG (N@K). The best results are in bold.}\label{tab:sota_msrs}
\setlength{\tabcolsep}{3pt}
\begin{tabular}{l|cc|cc}
\toprule
\multirow{2}{*}{Method}&\multirow{2}{*}{HR@10} & \multirow{2}{*}{NDCG@10} & \multirow{2}{*}{HR@20} & \multirow{2}{*}{NDCG@20} \\ 
\\
\midrule
\multicolumn{5}{c}{MicroLens-100K} \\
\midrule
HM4SR\cite{zhang2025hierarchical} (WWW'25)  & 0.0807 & 0.1207 & 0.0437 & 0.0538 \\
IISAN\cite{fu2024iisan} (SIGIR'24)  & 0.0960  & 0.1366 & 0.0526 & 0.0628 \\
\rowcolor{gray!20}
CROSSAN(T+I) & 0.0999 & 0.1428 & 0.0553 & 0.0661 \\
\rowcolor{gray!20}
CROSSAN (T+I+V+A)& \textbf{0.1033}  & \textbf{0.1452} & \textbf{0.0568} & \textbf{0.0673} \\
\midrule
\multicolumn{5}{c}{MicroLens-50K} \\
\midrule
HM4SR\cite{zhang2025hierarchical} (WWW'25)  & 0.0661 & 0.0976 & 0.0367 & 0.0446 \\
IISAN\cite{fu2024iisan} (SIGIR'24)  & 0.0771 & 0.1121 & 0.0421 & 0.0509 \\
\rowcolor{gray!20}
CROSSAN (T+I)     & 0.0806 & 0.1177 & 0.0431  & 0.0524 \\
\rowcolor{gray!20}
CROSSAN (T+I+V+A) & \textbf{0.0847} & \textbf{0.1222}  & \textbf{0.0462}  & \textbf{0.0557} \\
\bottomrule
\end{tabular}
\label{tab:microlens_combined}
% \vspace{-0.1in}
\end{table}

\begin{table}[h!]
\caption{Efficiency Comparison. TT, GM, and TP stand for Training Time (seconds/epoch), GPU Memory (MB), and Trainable Parameters, respectively. A lower value for each metric indicates an improvement in efficiency. We demonstrate the improvement over IISAN-E.}
\label{tab:eff_compare}
\renewcommand{\arraystretch}{1}
\renewcommand\tabcolsep{1.5pt}
\centering
\begin{tabular}{lccc}
\toprule
 \multirow{2}{*}{Method} &\multirow{2}{*}{TT ($\downarrow$)} & \multirow{2}{*}{GM ($\downarrow$)} & \multirow{2}{*}{TP ($\downarrow$)} \\ 
&&&\\
 \midrule
Full Finetuning (T+I) & 3,278 & 45,886 & 194,897,216\\ 
Adapter(T+I) & 2,856 & 35,652 & 38,017,088\\ 
LoRA(T+I) & 3,110 & 36,902 & 37,992,512\\ 
\midrule
\midrule 
IISAN-E$_{best}$ (T+I+V+A) & 213  & 5,556 & 58,432,824\\ 
IISAN-E$_{same}$ (T+I+V+A) & 207 & 4,476  & 30,889,784\\ 
\midrule
CROSSAN (T+I+V+A) \textbf{(ours) } & \textbf{144}  & \textbf{4,272}  & \textbf{24,741,456}\\ 
\midrule
Improvement$_{best}$& +32.39\% & +23.11\% & +57.66\% \\ 
Improvement$_{same}$ & +30.43\% & +4.56\% & +19.90\% \\ 
\bottomrule
\end{tabular}
\end{table}

\textbf{(Answer to RQ1)}: 
After extensive evaluation, we conclude that CROSSAN demonstrates superior scalability compared to the state-of-the-art efficient adaptation approaches. This is evidenced by its more substantial performance improvement when additional raw modalities are incorporated.

\subsection{Efficiency Evaluation (RQ2)}

In this section, we explore the efficiency of CROSSAN in terms of three dimensions: Training time, GPU memory, and Trainable Parameters, following the work by \cite{fu2024iisan}. We primarily focus on the \textbf{training-time} and \textbf{GPU Memory}, since they are the most important aspects of efficiency in practical settings. Due to computational limitations, we were only able to evaluate the efficiency of traditional adaptation approaches on image and text modalities\footnote{Adopting full fine-tuning or PEFT for the four encoders is nearly impossible given our GPU resources (we have access to only one A6000 GPU). 
Therefore, we leave this investigation for future work or for institutions with more extensive computational resources.}. 
Therefore, we primarily compare CROSSAN with IISAN-E (The intuitive extension of IISAN~\cite{fu2024iisan}), as other methods are too computationally intensive and not suitable for direct comparison with CROSSAN. In the following, we default to reporting efficiency based on the best performance.

\begin{table}[htbp]
\caption{Ablation Study on Fusion Method. D-Gated and S-Gated Fusion represents Dynamic Gated and Static Gated Fusion. Concat Fusion refers to the direct concatenation of all modalities.}
\label{tab:fusion}
\renewcommand{\arraystretch}{1.2}
\renewcommand\tabcolsep{2pt}
\centering
\begin{tabular}{clcccc}
\toprule
\multirow{2}{*}{Dataset}&
 \multirow{2}{*}{Method}& \multirow{2}{*}{HR@10} & \multirow{2}{*}{NDCG@10} & \multirow{2}{*}{HR@20} & \multirow{2}{*}{NDCG@20} \\ 
 \\
 \midrule
 \multirow{4}{*}{Microlens-100K}
&MOMEF & \textbf{0.1033}  & \textbf{0.0568} & \textbf{0.1452}  & \textbf{0.0673}  \\ 
&D-Gated Fusion & \underline{0.0994}   & \underline{0.0551} & \underline{0.1416}  & \underline{0.0657}  \\ 
&S-Gated Fusion & 0.0972   & 0.0532 & 0.1394  & 0.0638  \\
&Concat Fusion  & 0.0965   & 0.0531 & 0.1380  & 0.0636  \\ 
\midrule
 \multirow{4}{*}{Microlens-50K}
&MOMEF & \textbf{0.0847}  & \textbf{0.0462} & \textbf{0.1222}  & \textbf{0.0557}  \\ 
&D-Gated Fusion & 0.0796   & 0.0430 & \underline{0.1196}  & 0.0530  \\ 
&S-Gated Fusion & 0.0807   & 0.0437 & 0.1171  & 0.0529  \\
&Concat Fusion  & \underline{0.0816}   & \underline{0.0442} & 0.1181  & \underline{0.0534}  \\ 
\bottomrule
\end{tabular}
\end{table}

\begin{table}[h!]
\caption{Ablation Study on Cross- Vs. Independent-Modal. H and N represent the Hit Ratio and NDCG.}
\label{tab:abl_cross_modal}
\renewcommand{\arraystretch}{1}
\renewcommand\tabcolsep{3pt}
\centering
\begin{tabular}{clcccc}
\toprule
 \multirow{2}{*}{Dataset}& \multirow{2}{*}{Modality} & \multirow{2}{*}{HR@10} & \multirow{2}{*}{NDCG@10} & \multirow{2}{*}{HR@20} & \multirow{2}{*}{NDCG@20} \\ 
 \\
 \midrule
\multirow{2}{*}{Microlens-100K}&Cross-modal & \textbf{0.1033}  & \textbf{0.0568} & \textbf{0.1452}  & \textbf{0.0673}  \\ 
&Independent & 0.0993 & 0.0546 & 0.1418 & 0.0654  \\ 
\midrule
\multirow{2}{*}{Microlens-50K}&Cross-modal& \textbf{0.0847}  & \textbf{0.0462} & \textbf{0.1222}  & \textbf{0.0557}\\
&Independent&0.0815&0.0439&0.1169&0.0528\\
\bottomrule
\end{tabular}
\end{table}

As shown in \autoref{tab:eff_compare}, traditional adaptation methods with only two modalities reach the maximum of our available GPU memory. While IISAN-E reduces GPU memory, our proposed method, CROSSAN, is even more efficient and achieves better performance (see \autoref{tab:microlens_merged}). We attribute this observation to two key factors: (1) CROSSAN reduces the computational burden by using only one adapter tower per modality in each Multimodal Foundation Model (MFM), whereas IISAN-E requires an additional inter-modal adapter tower; (2) Hyperparameter tuning revealed that IISAN-E achieves its best performance with an embedding size of 1024, while CROSSAN reaches optimal performance with a hidden dimension of only 512, as shown in \autoref{fig:emb_independent_vs_cross}. This results in IISAN-E having more trainable parameters, which contributes to its reduced efficiency. Even when IISAN is configured with the same embedding dimension as CROSSAN, it remains less efficient due to the additional inter-modal adapter tower.  \textbf{(Answer to RQ2):} Although CROSSAN achieves superior performance, it also exceeds the efficiency of the state-of-the-art IISAN model.

\begin{figure*}[htbp]
  \centering
   \includegraphics[width=1\textwidth]{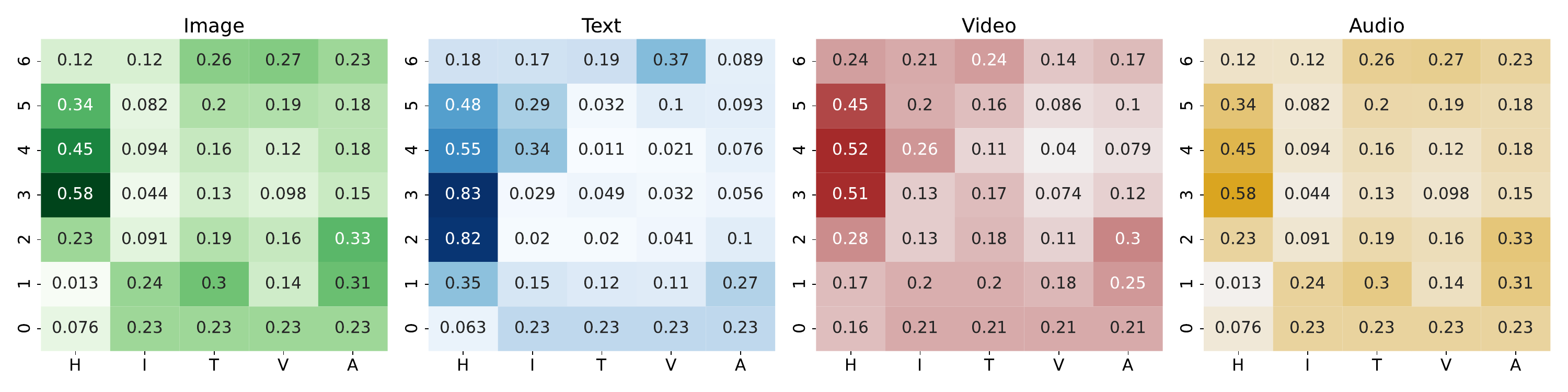}
   % \vspace{-0.1in}
  \caption{The heatmap visualization of CROSSAN's gates, where deeper colors indicate higher values. The y-axis corresponds to the layers of the side adapter, ranging from bottom (0) to top (6). ``H'' denotes the hidden state of the current layer's modality, while ``I'', ``T'', ``V'', and ``A'' represent the gate values for the previous side adapter layer's outputs corresponding to their respective modality.}
    \label{fig:gates_heatmap} 
    %\vspace{-0.1in}
\end{figure*}

\subsection{Ablation Study (RQ3)}

In this section, we present an ablation study focusing on two key components of CROSSAN: the fusion mechanism, MOMEF, and cross-modal interaction.

\begin{figure*}[htbp]
  \centering
   \includegraphics[width=\linewidth]{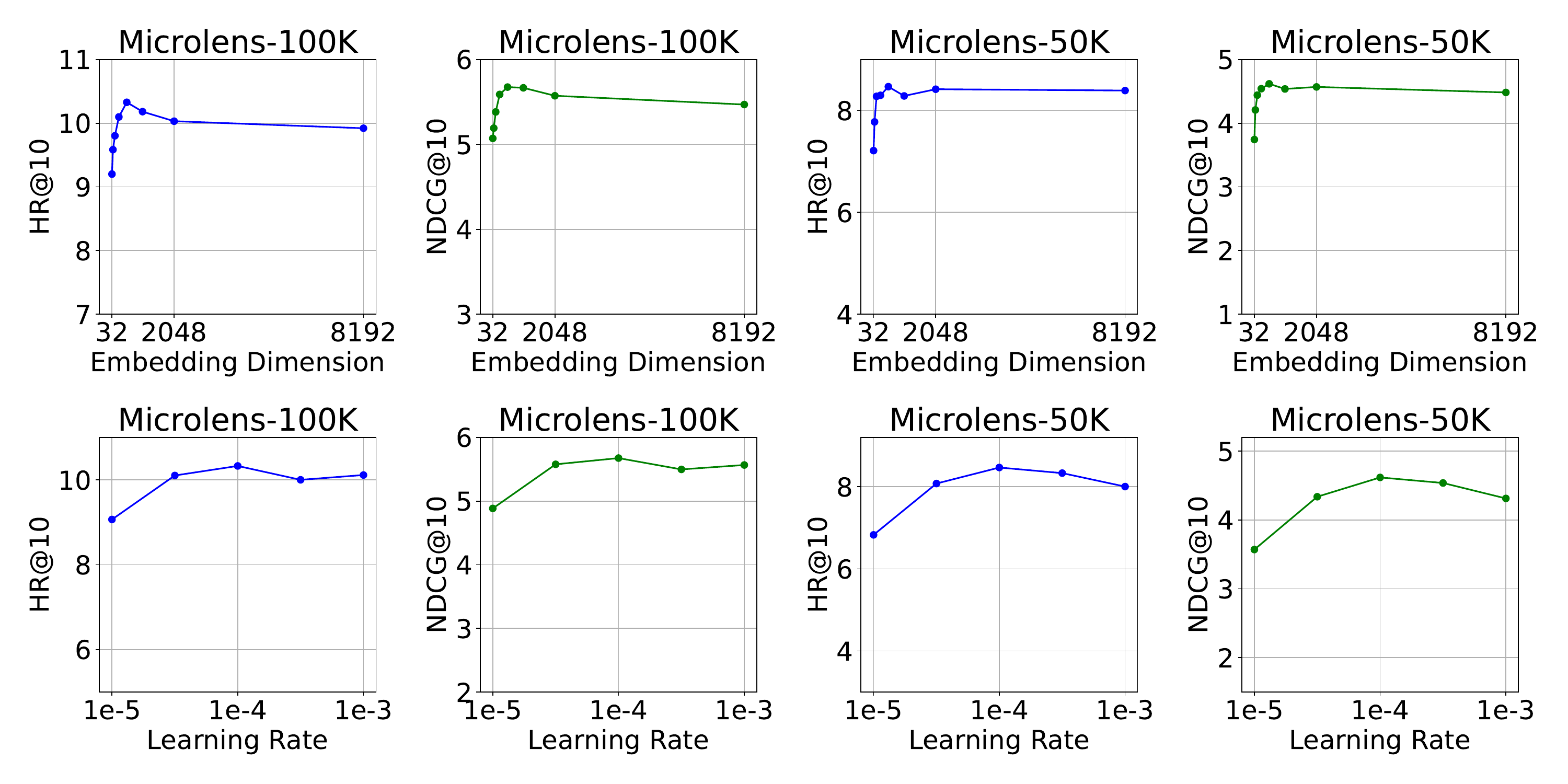}
  \caption{The first row of figures depicts CROSSAN's embedding dimension adaptation, where the embedding dimensions of the side adapters range from 32 to 8192.
The second row of figures illustrates CROSSAN's learning rate adaptation, with the learning rate varying between 1e-5 to 1e-3.}
    \label{fig:emb_lr_crossan} 
    % \vspace{-0.5em}
\end{figure*}

For the fusion mechanism, we compare MOMEF against three commonly used fusion methods: (1) Concat Fusion, which is widely adopted in existing literature~\cite{hu2021mmgcn,fu2024iisan,li2023multi}; (2) Static Gated Fusion (S-Gated), where a learnable gate is assigned to each modality and remains fixed after training, applying the same weights to all items; and (3) Dynamic Gated Fusion (D-Gated), which leverages a fully connected layer to generate different weights for each item, allowing more flexibility based on task-specific inputs. The proposed method, MOMEF, utilizes dynamic gating of input modalities to selectively activate the top-k modalities for each item. This approach offers a more fine-grained mechanism, allowing not only the weighting of modalities but also the precise selection of the most useful modalities.

As shown in \autoref{tab:fusion}, MOMEF outperforms all other methods, demonstrating its superior ability to fuse multiple modalities efficiently. 
While the D-gated method achieves the second-best results on the Microlens-100K dataset but it underperforms on the Microlens-50K dataset across three metrics when compared to direct concat. This suggests that other approaches may be less robust across different datasets.

Regarding cross-modal interaction, the results in \autoref{tab:abl_cross_modal} further confirm our preliminary findings (\autoref{sec:preliminary_study}): incorporating cross-modal interactions significantly enhances multimodal learning, validating its importance in achieving better performance. 
Additionally, we observed that the gating mechanism is crucial; without it, the model struggles to train properly, leading to a collapse in performance. We present the heatmap in \autoref{fig:gates_heatmap}, where we observe that the gate values for the hidden states in the corresponding modality tower are significantly larger in the middle layers, while the values in the lower layers are relatively smaller. This emphasizes the importance of the middle layers in MFMs. \textbf{(Answer to RQ3):} Both MOMEF and cross-modal interaction contribute to the overall performance of CROSSAN, with each approach demonstrating its usefulness.

\begin{table}[htbp]
\caption{Top-K experts for CROSSAN.}
\label{tab:experts}
\renewcommand\tabcolsep{3pt}
\centering
\begin{tabular}{cccccc}
\hline
 \multirow{2}{*}{Dataset}& \multirow{2}{*}{Top-K} & \multirow{2}{*}{HR@10} & \multirow{2}{*}{NDCG@10} & \multirow{2}{*}{HR@20} & \multirow{2}{*}{NDCG@20} \\ 
 \\
 \hline
\multirow{4}{*}{Microlens-100K}&4 &0.0992 & 0.0545 & 0.1422  & 0.0653 \\ 
&3 & 0.1002 & 0.0558&0.1459&0.0673  \\ 
&2 &\textbf{0.1033}  & \textbf{0.0568} & \textbf{0.1452}  & \textbf{0.0673}  \\ 
&1 & 0.1014  &0.0560 &0.1444&0.0668  \\ 
\hline
\multirow{4}{*}{Microlens-50K}&4 &0.0821 &0.0444 & 0.1189 &0.0536  \\ 
&3 & \textbf{0.0847}  & \textbf{0.0462} & \textbf{0.1222}  & \textbf{0.0557}  \\ 
&2 & 0.0815 &0.0438 & 0.1197&0.0534  \\ 
&1 & 0.0822 & 0.0451&0.1178&0.0541  \\ 
\hline
\end{tabular}
\end{table}

\subsection{Hyperparameter Analysis (RQ4)}

In this section, we explore three key hyperparameters: (1) learning rate, (2) hidden dimension, and (3) number of experts. The former two are fundamental parameters commonly explored in adapter-based recommendation models \cite{fu2024exploring}, while the number of experts is introduced by MOMEF. To optimize the model’s performance, we conduct an extensive hyperparameter search for these settings.

\textbf{Embedding Dimension.} As shown in \autoref{fig:emb_lr_crossan}, the performance of CROSSAN demonstrates a clear dependency on the embedding dimension size. A relatively large embedding dimension is essential for effective adaptation. In contrast, smaller embedding leads to noticeable drops in performance. However, due to the efficiency of CROSSAN’s adapter-based architecture, increasing the embedding dimension does not result in significant computational overhead, such as extended training time or excessive GPU memory usage. Notably, when scaling the embedding size up to 8192, performance remains stable, suggesting that CROSSAN maintains its effectiveness as long as the embedding size exceeds a certain threshold.

\textbf{Learning Rate.} \autoref{fig:emb_lr_crossan} illustrates the effect of learning rate on model performance. CROSSAN's performance appears to remain stable once the learning rate exceeds 5e-5. However, using a smaller learning rate (e.g., 1e-5) results in suboptimal performance. This highlights the necessity of fine-tuning this hyperparameter within a suitable range to achieve optimal performance.

\textbf{Top-K experts.} The Top-K is a new hyperparameter introduced through MOMEF. As shown in \autoref{tab:experts}, the optimal number of experts differs across datasets. For example, the Microlens-100K dataset performs best with two experts, while three experts yield the best results for the Microlens-50K dataset. These findings indicate that both very large and small numbers of experts are ineffective.

\textbf{(Answer to RQ4):} Based on the experiments, we conclude two observations from the hyperparameter analysis: (1) The embedding dimension and learning rate for CROSSAN should be set within a large range to ensure stable performance. (2) Top-K experts are dataset-specific and typically lie within a moderate range.

\begin{table*}[!t]
\caption{Performance (\%) comparison across methods on three datasets. Best results in bold, second-best underlined.}
\centering
\label{tab:am_crossan}
\begin{tabular}{c|cc|cc|cc}
\hline
\multirow{2}{*}{Method} & \multicolumn{2}{c|}{Scientific} & \multicolumn{2}{c|}{Instrument} & \multicolumn{2}{c}{Office} \\
& HR@10 & NDCG@10 & HR@10 & NDCG@10 & HR@10 & NDCG@10 \\
\hline
FFT         & 6.62 & 4.06 & 8.76 & 6.76 & 6.30 & 4.58 \\
Adapter      & 6.61 & 3.91 & 8.82 & 6.83 & 6.65 & 4.85 \\
LoRA         & 6.62 & 4.09 & 8.43 & 6.64 & 6.55 & 4.78 \\
BitFit       & 6.37 & 3.76 & 8.65 & 6.71 & 6.78 & 4.87 \\
IISAN        & \underline{6.83} & \underline{4.14} & \underline{9.06} & \underline{7.01} & \underline{6.80} & \underline{4.92} \\
\rowcolor{gray!20}
\textbf{CROSSAN} & \textbf{6.90} & \textbf{4.22} & \textbf{9.08} & \textbf{7.05} & \textbf{6.83} & \textbf{4.95} \\
\hline
\end{tabular}
\end{table*}

\begin{table}[!t]
\centering
\caption{Performance comparison on MicroLens-100K datasets. The best results are in bold. The second best is underlined.}
\label{tab:llm}
\begin{tabular}{l|cc|cc}
\toprule
\multirow{2}{*}{Method}                         & \multirow{2}{*}{H@10}   & \multirow{2}{*}{N@10}   & \multirow{2}{*}{H@20}   & \multirow{2}{*}{N@20}   \\ 
\\
\midrule
CROSSAN & 0.1033 & 0.1452 & \underline{0.0568}  & 0.0673 \\
\hline
CROSSAN (QWen-7B)   & 0.1033  & \textbf{0.1464} & 0.0564 & \underline{0.0674} \\
CROSSAN (QWen-VL-7B)   & 0.1033  & \underline{0.1460} & \textbf{0.0570} & \textbf{0.0677} \\
\midrule
\end{tabular}
\end{table}

\begin{table}[ht]
\centering
\renewcommand\tabcolsep{3pt}
\caption{Performance Comparison of Independent vs. Cross-Modal Architectures with Cross-Attention Fusion on the MicroLens-100K Dataset.}
\label{tab:cross-attn}
\begin{tabular}{lcccc}
\hline
\multirow{2}{*}{Modality}                         & \multirow{2}{*}{H@10}   & \multirow{2}{*}{N@10}   & \multirow{2}{*}{H@20}   & \multirow{2}{*}{N@20}   \\ 
\\
\hline
Independent (w/ MOMEF Fusion)                     & 0.0993 & 0.0546 & 0.1418 & 0.0654 \\
Independent (w/ Cross-attn Fusion) & 0.1008 & 0.0558 & 0.1423 & 0.0663 \\
\hline
Cross-modal (w/ MOMEF Fusion)                     & 0.1033 & 0.0568 & 0.1452 & 0.0673 \\
Cross-modal (w/ Cross-attn Fusion) & 0.1014 & 0.0561 & 0.1435 & 0.0668 \\ \hline
\end{tabular}
\end{table}

\section{More evaluations}
\label{sec:ap_more_eval}
\noindent \textbf{Exploration of More Datasets.} To further investigate the effectiveness of CROSSAN, we conducted experiments on three Amazon datasets used in \cite{fu2024iisan}. We compared the performance of CROSSAN against full fine-tuning and various PEFT methods, including Adapter, LoRA, BitFit, and IISAN. Our results show that CROSSAN achieves the best adaptation performance across all three datasets using only two modalities as described in \autoref{tab:am_crossan}. Although the improvement over IISAN on these datasets is not on a large margin, CROSSAN offers greater efficiency by requiring only two side adapter towers, compared to the three needed by IISAN. Consequently, even with just two modalities, CROSSAN proves to be an optimal solution for these datasets.

\noindent \textbf{Exploration of LLM and LVM.} To assess the impact of incorporating a recent large language model (LLM)\footnote{\url{https://huggingface.co/Qwen/Qwen2-7B}} and large vision-language model (LVM)\footnote{\url{https://huggingface.co/Qwen/Qwen2-VL-7B}}, we follow the implementation of \cite{fu2024efficient} by replacing the original text tower. This substitution yields a slight performance improvement for CROSSAN when using Qwen-VL-7B, while Qwen-7B performs marginally worse in terms of NDCG@10 and marginally improves the other metrics (\autoref{tab:llm}). Nonetheless, while the implementation of LLMs can provide slight performance gains, the simple BERT-based text tower used in our main paper already delivers strong and competitive results.

\noindent \textbf{More Investigation on Independent Variant.} In some cases, the independent side-adapter variant still has merit, as it avoids layer-wise cross-modal operations and may be more suitable for asymmetrical architectures~\cite{fu2024efficient}. Understanding why it underperforms compared to the cross-modal approach is therefore meaningful. A natural hypothesis is that the independent variant underperforms because it lacks sufficient interaction during layer-wise fusion, and the sparse fusion in MOMEF may not fully realize its potential. To investigate this, we first examine the effect of varying the number of experts to mitigate the sparsity factor, as summarized in \autoref{tab:ind_experts}. The results indicate that the independent approach requires a relatively larger number of experts in MOMEF. In contrast, the cross-modal method reduces the optimal number of experts by promoting a more unified feature representation, thanks to its higher mutual information.

To further enhance the final fusion interaction, we adopt a commonly used cross-attention fusion module to the independent variant. Unlike MOMEF, this module performs dense computation (see \autoref{tab:cross-attn}). The results support our hypothesis: cross-attention yields small improvements over MOMEF fusion in the independent variant (e.g., HR@10 improves from 0.0993 to 0.1008), aligning with the intuition that cross-attention can help in this setting. However, when integrated into CROSSAN’s interaction layers, performance slightly declined (e.g., Hit@10: 0.1033 → 0.1014), suggesting that cross-attention is not always advantageous when strong cross-modal interactions are already present. These findings may provide useful insights for future work that adopts the independent variant, highlighting that cross-attention can be beneficial compared with MOMEF.

\section{Related Work}
\noindent \textbf{Multimodal Foundation Models (MFMs).}
Recent advances in multimodal learning leverage pre-trained models to enhance downstream task performance while reducing pre-training costs \cite{sun2023generative, chen2024knowledge, zhang2024tamm,he2025double,ge2024towards}. BERT \cite{devlin2018bert} pioneered the pretraining and fine-tuning paradigm in NLP, while Vision Transformer (ViT) \cite{dosovitskiy2020image} adapted this approach for image classification. CLIP \cite{radford2021learning} bridged vision and language through contrastive learning, enabling robust zero-shot capabilities. Large-scale models like GPT-4 \cite{achiam2023gpt}, T5 \cite{raffel2020exploring}, and multimodal variants like DALL-E \cite{ramesh2021zero} and Flamingo \cite{alayrac2022flamingo} expanded the ability to process diverse modalities. In video representation learning, models such as SlowFast-R50 \cite{feichtenhofer2019slowfast}, MViT-b \cite{fan2021multiscale}, and VideoMAE \cite{tong2022videomae} effectively capture both temporal and spatial features. Meanwhile, audio models like Wave2Vec \cite{baevski2020wav2vec} and the Audio Spectrogram Transformer (AST) \cite{gong2021ast} enhance audio classification by operating on spectrograms. Together, these models highlight the increasing strength of multimodal learning across a variety of tasks.

\begin{table}[htbp]
\caption{Performance comparison of Independent and Cross-modal methods.}
\label{tab:ind_experts}
\renewcommand\tabcolsep{5pt}
\centering
\begin{tabular}{cccccc}
\hline
\multirow{2}{*}{Method} & \multirow{2}{*}{Top-K} & \multirow{2}{*}{H@10} & \multirow{2}{*}{N@10} & \multirow{2}{*}{H@20} & \multirow{2}{*}{N@20} \\ 
\\
\hline
\multirow{4}{*}{Independent} 
&1 & 0.0973 & 0.0535 & 0.1404 & 0.0643 \\ 
&2 & \textbf{0.0993} & \textbf{0.0546} & 0.1418 & 0.0654 \\ 
&3 & \textbf{0.0993} & \textbf{0.0546} & \textbf{0.1432} & 0.0657 \\ 
&4 & 0.0989 & 0.0540 & 0.1411 & 0.0646 \\ 
\hline
\multirow{4}{*}{Cross-modal} 
&1 & 0.1014 & 0.0560 & 0.1444 & 0.0668 \\ 
&2 & \textbf{0.1033} & \textbf{0.0568} & 0.1452 & \textbf{0.0673} \\ 
&3 & 0.1002 & 0.0558 & \textbf{0.1459} & \textbf{0.0673} \\ 
&4 & 0.0992 & 0.0545 & 0.1422 & 0.0653 \\ 
\hline
\end{tabular}
\end{table}

\noindent \textbf{Efficient Adaptation of MFMs in RS.}
The RS field has progressively investigated the incorporation of diverse modalities to improve the effectiveness of recommendations \cite{zhang2024multimodal,ye2025harnessing,zhou2023mmrec,hu2023adaptive,liu2024multimodal,tang2023one,wu2021mm,sun2020multi,yuan2023go,wei2024fummer,li2023exploring,wu2024promise,wei2023multi,wang2023missrec,cheng2023image,wei2024fummer,liu2023id,qu2023thoroughly,hu2024lightweight,liu2024once,bao2023tallrec,zhang2023beyond,huan2024exploring,wangautomated,lin2023can,khan2025exploiting,zhou2025large,qu2025efficient,fu20251st}. Recent studies \cite{elsayed2022end,yuan2023go,ni2023content,li2023multi} have demonstrated the superiority of the MoRec framework using end-to-end learning, showing it significantly outperforms traditional approaches that rely on offline feature extraction. For instance, \cite{li2023multi} and \cite{fu2024iisan} highlighted the advantages of end-to-end training, which jointly leverages both image and text modalities, compared to methods that employ a single modality. Despite the strong performance of raw content learning, a major drawback of these approaches is the continued dependence on full fine-tuning of large multimodal encoders, leading to performance inefficiencies. 

Parameter-efficient fine-tuning (PEFT) methods have made strides in addressing this issue, as shown in works like M6-Rec~\cite{cui2022m6}, Tallrec~\cite{bao2023tallrec}, and AdapterRec~\cite{fu2024exploring}, which demonstrate that PEFT techniques can achieve competitive performance with significantly reduced overhead. However, many PEFT methods still rely on established approaches, often overlooking practical efficiency concerns.

Much of the existing research predominantly focuses on traditional Adapter or LoRA-based solutions and is limited to single-modality adaptation due to the inefficiency of these adaptation approaches. A recent study, IISAN \cite{fu2024iisan}, introduced a structure that utilizes independent adapters within each tower, complemented by a single inter-adapter for image and text adaptation. However, the inter-tower has a fixed input and lacks sufficient cross-modal interaction. In contrast, CROSSAN emphasizes cross-modal side adapters with joint learning and updates. Additionally, IISAN considers only image and text scenarios, leaving many other existing modalities unexplored. Upon attempting to expand this method, we concluded that it lacked the desirable scalability for additional modalities. To the best of our knowledge, cross-modality side adapters are largely underrepresented in the existing literature on recommender systems. Furthermore, CROSSAN investigates the novel area of scalable and efficient adaptation for more ($>$2) MFMs with raw modalities, which will facilitate future research.

\section{Conclusion}
CROSSAN provides a scalable, efficient, and effective approach for adapting multiple multimodal foundation models (MFMs) in sequential recommendation tasks. By incorporating cross-modal side adapters along with the Mixture of Modality Expert Fusion (MOMEF), CROSSAN achieves superior performance. Extensive experimental results validate the approach's ability to improve recommendation effectiveness as additional modalities are integrated, demonstrating its superiority over existing methods.

Future research can extend CROSSAN to various multimodal tasks, including multimodal classification \cite{gomez2015multimodal}, retrieval \cite{rafailidis2013unified}, and generative modeling \cite{dai2022enabling}, where the integration of diverse data modalities is critical. These directions offer significant potential to improve both the efficiency and performance of multimodal learning across a broad range of applications, positioning CROSSAN as a general adaptation approach for future research.

\bibliographystyle{IEEEtran}
\bibliography{cas-refs}

\begin{IEEEbiography}[{\includegraphics[width=1in,height=1.25in,clip,keepaspectratio]{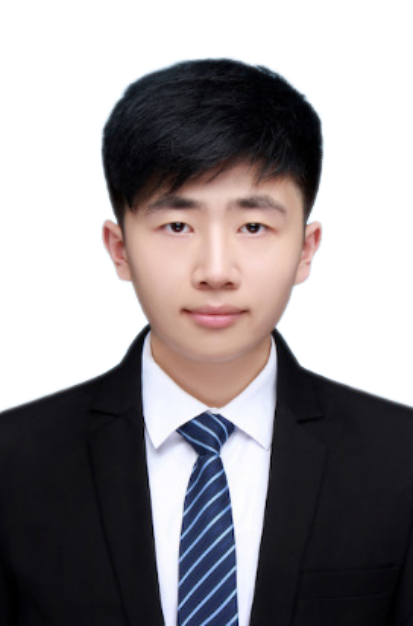}}]{Junchen Fu}
is a third-year PhD candidate under the supervision of Prof. Joemon Jose at the School of Computing Science, University of Glasgow. He has contributed to several leading AI conferences and journals, including ICML, SIGIR, WWW, WSDM, CIKM, MM,  CVPR, and TPAMI. He also serves as the program committee member and invited reviewer for premier conferences and journals, such as ICLR, SIGIR, KDD, WSDM, CIKM, AAAI, MM, TKDE, TOIS, TMM, and TORS. His research primarily focuses on enhancing the efficiency of adapting multimodal foundation models for recommendation. 
\end{IEEEbiography}
% \vskip -0.4in
\begin{IEEEbiography}[{\includegraphics[width=1in,height=1.25in,clip,keepaspectratio]{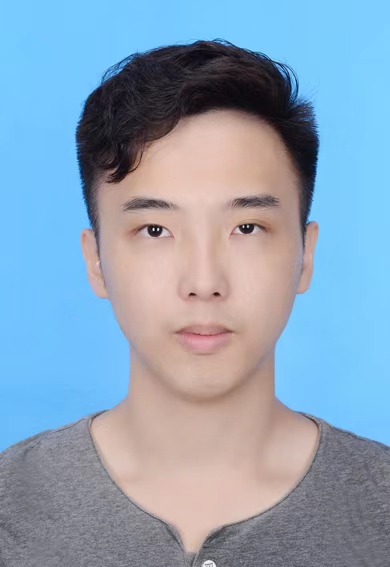}}]{Yongxin Ni} is currently a PhD student at Boston University. He received the B.E. degree in Computer Science and Technology from Shenzhen University, the M.Tech. degree in Intelligent Systems from National University of Singapore.  He has contributed to several leading AI conferences and journals, including CVPR, SIGIR, WWW, WSDM, CIKM, AAAI, and TPAMI. He also serves as reviewers for top-tier conferences and journals, such as ICLR, SIGIR, WSDM, CIKM, AAAI, MM, TKDE, TOIS, TMM, TORS, IP\&M, and TIP. His research interests include deep learning, multi-modality representation and recommender systems. 
\end{IEEEbiography}

% \vskip -0.5in
\begin{IEEEbiography}[{\includegraphics[width=1in,height=1.25in,clip,keepaspectratio]{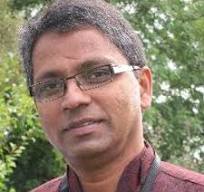}}]{Joemon Jose} is a Professor with the School of Computing Science, University of Glasgow,
Glasgow, and a member of the Information Retrieval
Group. He has published over 300 papers with more
than 10,000 Google Scholar citations, and an H-index
of 51. He leads the Multimedia Information Retrieval
group which investigates research issues related
to the above themes. His research focuses around
the following three themes: social media analytics,
multimodal interaction for information retrieval, and
multimedia mining and search.
\end{IEEEbiography}

% \vskip -0.5in
\begin{IEEEbiography}[{\includegraphics[width=1in,height=1.25in,clip,keepaspectratio]{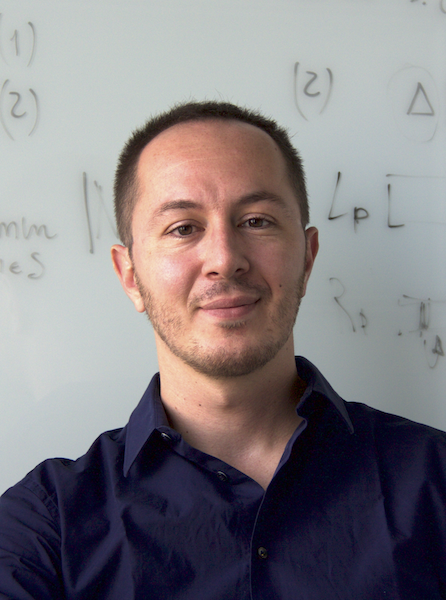}}]{Ioannis Arapakis} is a Principal Research Scientist at Telefónica Research, specializing in behavior interpretation algorithms for user modeling in both offline and online contexts, with a focus on web search. He holds a Ph.D. in Information Retrieval from the University of Glasgow, and an M.Sc. in Information Technology from the Royal Institute of Technology. Previously, he was a Research Scientist at Yahoo Labs, where he worked on data mining, IR, HCI, and multimedia mining projects. 
\end{IEEEbiography}
% \vskip -0.9in
\begin{IEEEbiography}[{\includegraphics[width=1in,height=1.25in,clip,keepaspectratio]{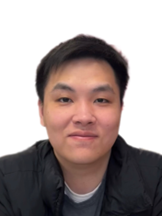}}]{Kaiwen Zheng} is a third-year Ph.D. candidate in the School of Computing Science at the University of Glasgow, under the supervision of Prof. Joemon Jose. His research primarily focuses on multimodal expression recognition. He has published in leading AI conferences and journals, including ICML, WWW, and ICME, and serves as an invited reviewer for premier venues such as AAAI, IEEE Transactions on Affective Computing (TAFFC), ICME, and CIKM.
\end{IEEEbiography}

% \vskip -0.9in
\begin{IEEEbiography}[{\includegraphics[width=1in,height=1.25in,clip,keepaspectratio]{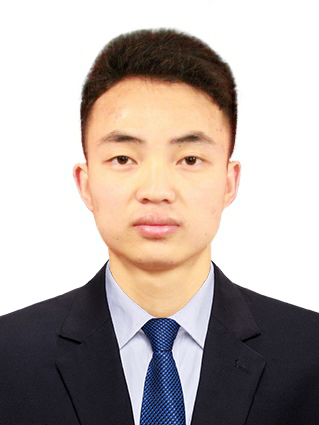
}}]{Youhua Li} is currently pursuing his Ph.D. in Decision Analytics and Operations at the City University of Hong Kong. He received his M.S. degree in Computer Science and Technology from ShanghaiTech University, and his B.S. degree in Automation Engineering from the University of Electronic Science and Technology of China. He has published in top-tier conferences and journals such as CVPR, TPAMI, ICDE,WWW, WSDM, and SIGIR. His research interests span data-driven decision making and operations optimization, large language models (LLMs) and agents.
\end{IEEEbiography}

% \vskip -0.5in
\begin{IEEEbiography}[{\includegraphics[width=1in,height=1.25in,clip,keepaspectratio]{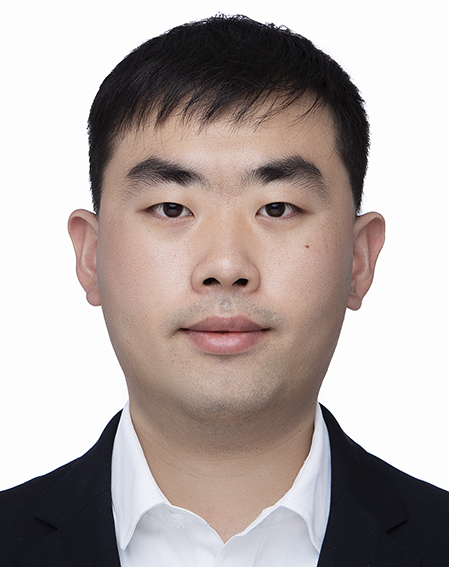}}]{Xuri Ge} is now a tenure-track assistant professor in the school of artificial intelligence, Shandong University. He earned his PhD at the University of Glasgow (UK) and received M.S. degree from Xiamen University (China). His current research interests include computer vision, multimodal representation learning, and information retrieval. He has contributed to several leading conferences and journals, including NeurIPS, ICML,  SIGIR, ACM MM, CIKM, WSDM, ACM TIST, and IP\&M, etc. He serves as the PC member and reviewer for conferences and journals, such as NeurIPS, MM, ICLR, TKDE, TOIS, IJCV, etc.
\end{IEEEbiography}

\end{document}